\documentclass[a4paper]{jpconf}
\usepackage{graphicx}
\pdfoutput=1
\begin{document}
\title{Doping evolution of the electronic specific heat coefficient 
in slightly-doped La$_{2-x}$Sr$_x$CuO$_4$ single crystals}

\author{Seiki Komiya and I Tsukada}

\address{Central Research Institute of Electric Power Industry, Yokosuka, Kanagawa 240-0196, Japan}

\ead{komiya@criepi.denken.or.jp}

\begin{abstract}

Detailed doping dependence of the electronic specific heat coefficient $\gamma$ is 
studied for La$_{2-x}$Sr$_x$CuO$_4$ (LSCO) single crystals in the slightly-doped 
regime. 
We find that $\gamma$ systematically increases with doping, and furthermore, 
even for the samples in the antiferromagnetic (AF) regime, $\gamma$ already acquires finite 
value and grows with $x$. This suggests that finite electronic density of states (DOS) 
is created in the AF regime where the transport shows strong localization at low temperatures, 
and this means the system is not a real insulator with a clear gap even though it still keeps 
long range AF order. 

\end{abstract}

\section{Introduction}

It is still an issue of controversy in the study of high-$T_c$ cuprates how carriers are 
doped into the parent Mott insulator and the metallic state arises with doping. 
Transport measurements have demonstrated that 
metallic in-plane resistivity is immediately realized at high temperatures 
with 1\%-hole doping, although 
it turns insulating at low temperatures\cite{Ando_PRL01}. 
This behavior is qualitatively the same even in the underdoped superconducting regime 
when superconductivity is suppressed by magnetic field or Zn doping\cite{Boebinger_PRL96, 
Ono_PRL00, Komiya_PRB04}. 
High temperature Hall coefficient ($R_H$) measurements 
have shown that the temperature 
dependences of $R_H$ in the LSCO system can be fitted as thermal activation type 
at high temperatures\cite{Ono_PRB07}, suggesting that the carrier concentration 
would be temperature dependent. 
The background electronic structure of this peculiar transport 
is extensively studied by angle resolved photoemission spectroscopy (ARPES) 
experiments, and some kind of fragmented Fermi surface, ``Fermi arc", is found 
to grow with doping\cite{Teppei_PRL, Teppei_JPCM}. 
However, the identity of this novel electronic state is not fully understood yet. 
There is also a long debate on the metallicity of the doped Mott insulators\cite{Nagaosa_RMP, Phillips_PRL05}. 

To elucidate the doping evolution of the electronic states 
in a different light, we study the detailed doping 
dependence of the electronic specific heat coefficient $\gamma$ of LSCO 
in slightly doped regime. We find that for the parent insulating La$_2$CuO$_4$ 
(LCO), $\gamma$ is indeed zero, but it systematically increases with 
doping even though the system keeps long range antiferromagnetic order 
($0.005 \leq x < 0.02$). This result means that the finite electronic density of 
states is created 
already with 0.5\%-hole/Cu doping.

\section{Experimental}

Single crystals of LSCO are grown by traveling solvent floating zone method\cite{Komiya_PRB02}. 
The nominal Sr concentrations of the measured crystals are 0, 0.005, 0.01, 0.015, 
0.02, 0.03, 0.04, and 0.05. 
These LSCO crystals in slightly-doped regime tend to have excess oxygen that leads the samples 
showing minor superconductivity, so these are annealed in pure Ar to remove 
excess oxygen. Samples with $x = 0$ to 0.015 show N\'{e}el transition and $T_N$s of 
these samples are determined by susceptibility measurement  
to be 320 K, 280 K, 240K, and 200 K for $x = 0$, 0.005, 0.01, and 0.015, 
respectively. Samples with typical weight of 20 mg are used for heat capacity 
measurement which is performed by the relaxation method, 
using Quantum Design's Physical Properties Measurement System down to 2 K. 
Transport properties are measured by the conventional four-terminal method. 

\section{Results and discussions}

In Fig. 1, we show the $C_p/T$ vs. $T^2$ plot of measured samples at low temperatures. 
We see that the low temperature specific heat gradually and systematically 
increases with doping, and 
for all samples studied here, $C_p/T$ has almost linear dependence on $T^2$ and 
Schottky anomaly is hardly observed in this temperature range. Thus we can reasonably 
fit the data in a form of $\gamma T + \beta T^3$. 
Note that the transition to the Mott insulating state in high-$T_c$ cuprates is 
not of the diverging effective mass type\cite{Kumagai_PRB93, Padilla_PRB05}, 
and the $\gamma$ values are not enhanced when the system approaches 
the parent Mott insulator. 

\begin{figure}[b]
\begin{minipage}{18pc}
\includegraphics[width=18pc]{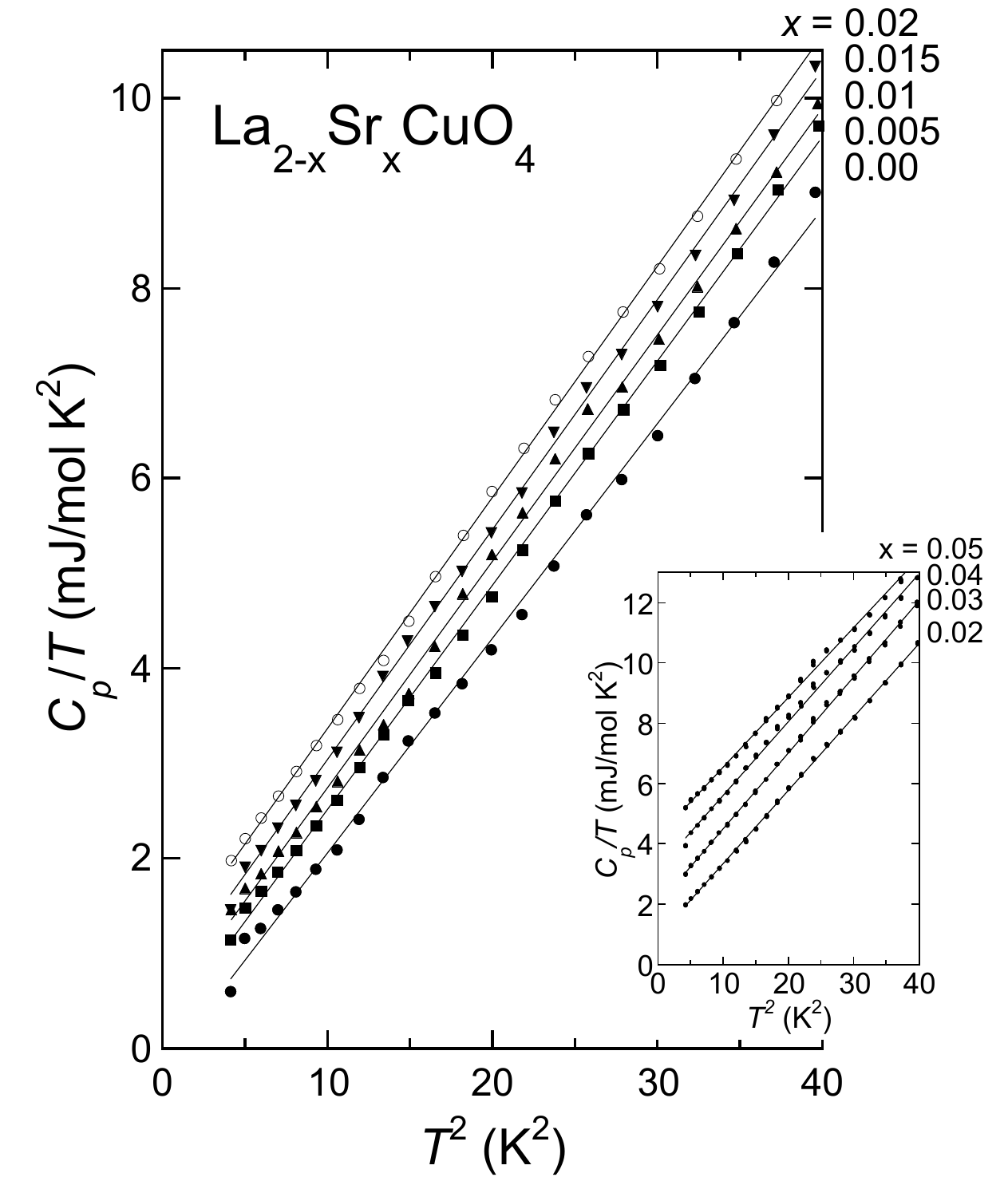}
\caption{\label{label}$C_p/T$ vs. $T^2$ plot for slightly doped LSCO crystals. 
Data for $0 \leq x \leq 0.02$ and $0.02 \leq x \leq 0.05$ are separately shown in the main panel and the inset, respectively.}
\end{minipage}\hspace{2pc}%
\begin{minipage}{18pc}
\includegraphics[width=18pc]{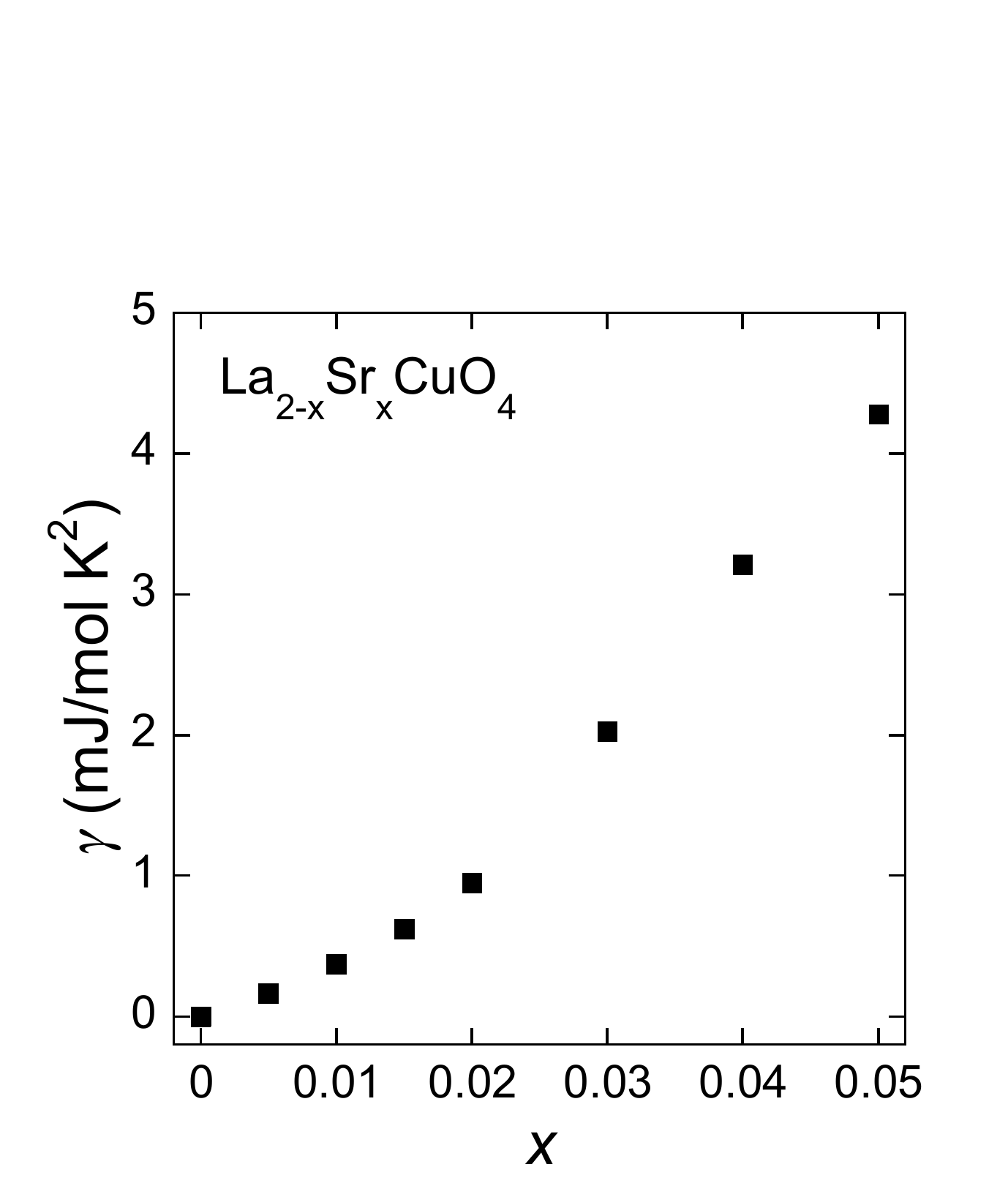}
\caption{\label{label}Doping dependence of the electronic specific heat coefficient.}
\end{minipage} 
\end{figure}

The doping dependence of $\gamma$ is presented in Fig. 2. 
For the parent LCO, $\gamma$ is zero within an experimental error. 
This result is consistent with the high temperature Hall measurement. 
The temperature dependence 
of $R_H$ for the parent LCO is found to be well 
fitted as a thermal activation type with two different 
activation energies\cite{Ono_PRB07}, so $\gamma = 0$ is 
reasonable as an insulator with a definitive gap. When Sr is doped, 
$\gamma$ immediately grows even in the AF ``insulating" regime. 
This is not consistent 
with ARPES results which suggest a full gap in the AF regime\cite{Kyle_Shen_PRB}. 
It was also pointed out, however, that the length of the ``arc'' is in good correlation 
with the $\gamma$ value in a wide doping range\cite{Teppei_JPCM, Momono_PhysicaC}, 
and therefore similar arc with finite length which is too small to be detected by ARPES measurements 
would exist in the AF regime. 

In Fig. 2, it is also seen that the doping dependence of $\gamma$ changes at $x=0.02$ where 
the long range AF order disappears. To see this difference in more detail, we compare 
the behavior of low temperature resistivity across $x=0.02$. As presented in Figs. 3 and 4, 
$\rho_{ab}$ for the $x=0.03$ crystal shows well-known variable range hopping behavior, 
while for the $x=0.015$ sample, 
low temperature resistivity can be best fitted as $exp(T^{-1/2})$ in a rather wide temperature range. 
This temperature dependence suggests that a gap, possibly a soft Coulomb gap\cite{Efros}, 
is opening with decreasing temperature 
for LSCO with $x=0.015$. This behavior is reasonable because the effective Coulomb interaction 
increases when the number of carriers decreases. DOS at the Fermi level would be suppressed 
by this gap, causing the $x$ dependence of $\gamma$ weaker in the AF regime. 
Note that the finite $\gamma$ value suggests that this gap may not fully 
open and DOS still remains. ARPES experiments would detect this gap 
opening.

\begin{figure}[t]
\begin{minipage}{18pc}
\includegraphics[width=18pc]{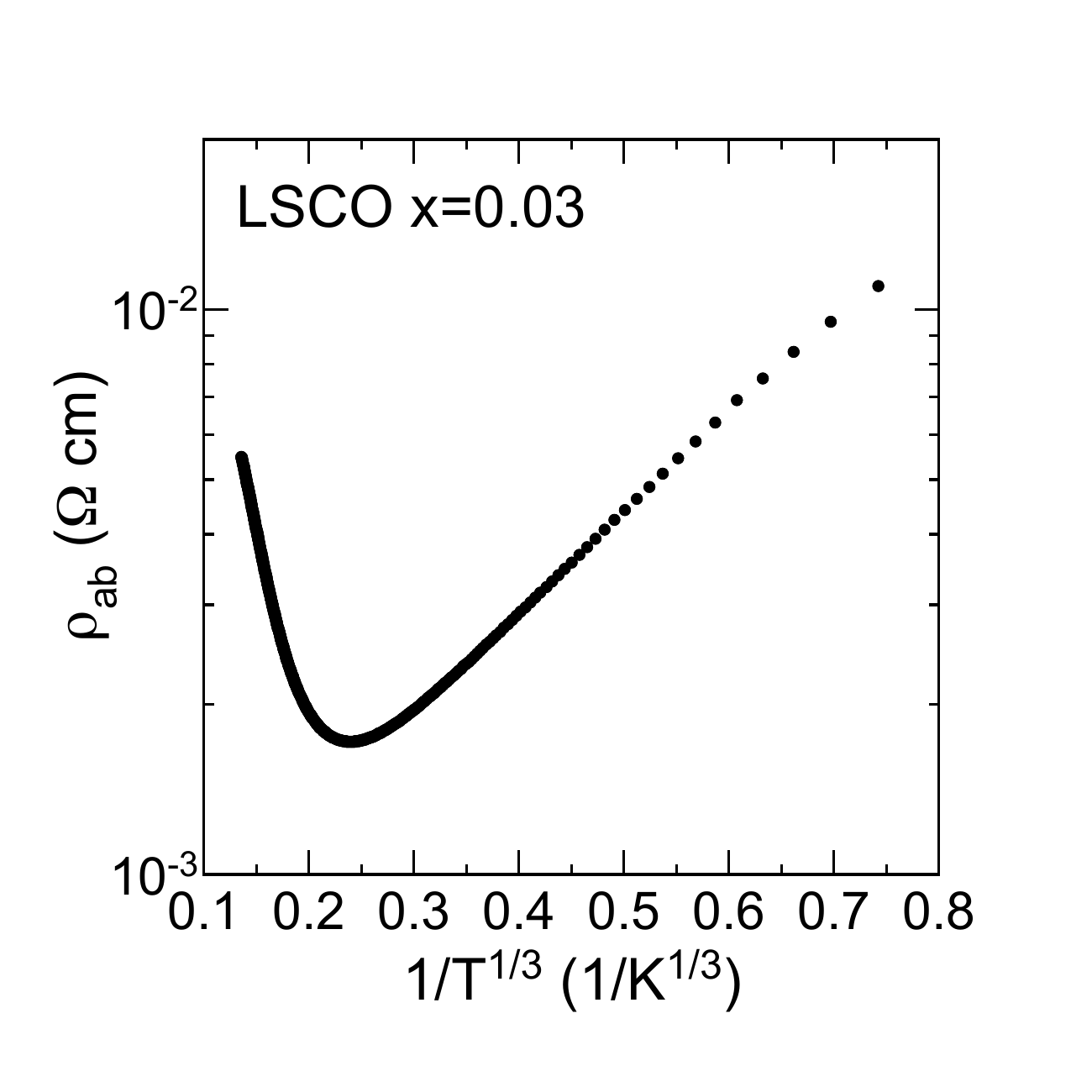}
\caption{\label{label}Variable range hopping behavior in $\rho_{ab}$ for LSCO with $x=0.03$.}
\end{minipage}\hspace{2pc}%
\begin{minipage}{18pc}
\includegraphics[width=18pc]{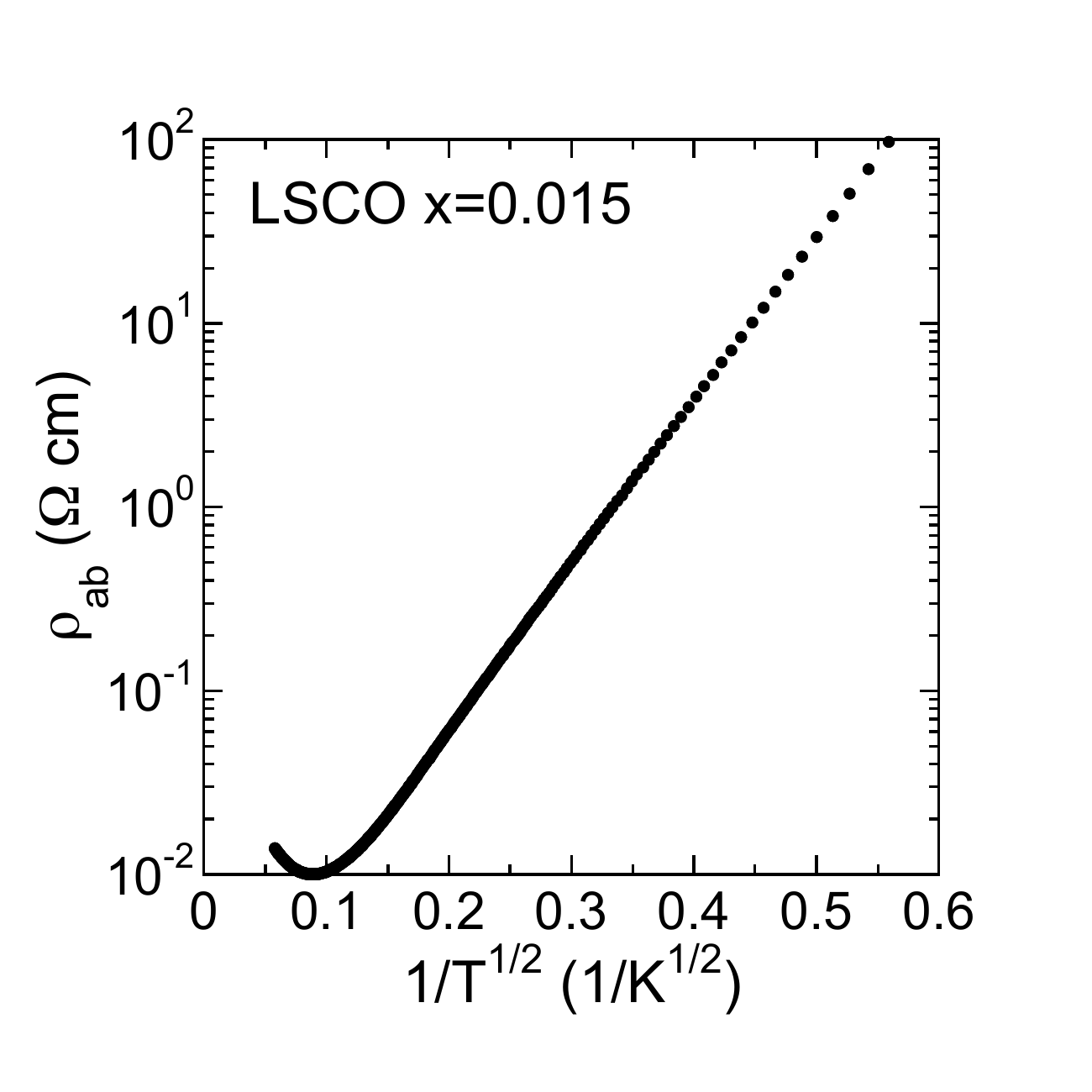}
\caption{\label{label}$\rho_{ab}$ for LSCO with $x=0.015$. Coulomb-gap like temperature dependence is observed below $\sim$ 10 K. }
\end{minipage} 
\end{figure}

How do the finite DOS and the AF order coexist at $0.005\leq x < 0.02$ ? 
One possible scenario is related to the domain structure observed in neutron 
scattering experiments and the spin glass behavior at low temperatures 
in the AF regime\cite{Matsuda_PRB02, Niedermayer_PRL98}. 
They have found that in the AF LSCO samples with $0.01 \leq x < 0.02$, some part of 
the AF ordered phase turns into a spin glass phase 
at low temperatures. The observed $\gamma$ in the AF regime would come 
from this spin glass portion in the background AF order. 
Magnetotransport experiments support this domain-structure scenario, where 
large negative magnetoresistance suddenly appears above a magnetic field of the weak ferromagnetic 
transition\cite{Ando_PRL03}. 

Although the transport properties show the strong localization in 
slightly to underdoped regime at low temperatures\cite{Ando_PRL01, Boebinger_PRL96}, 
we emphasize that the system is not an insulator with a gap in the energy spectrum. 
This can be also understood by comparing the behavior of the thermal conductivity and 
the heat capacity. In LSCO, low temperature thermal conductivity measurements have 
revealed that the residual electronic term of the thermal conductivity at zero temperature limit 
is zero in the non-superconducting samples with $0 \leq x \leq 0.05$\cite{Takeya_PRL02}, 
while the electronic 
specific heat coefficient increases monotonically with $x$ in the same doping range. 
Therefore, the observed resistivity divergence is simply because the mean free length of carriers 
decreases with decreasing temperature. Although the localization mechanism itself is still 
unclear, the electronic ground state may not be different regardless of whether the 
long range AF order is established. 

\section{Summary}
Low temperature electronic specific heat is studied for the LSCO system in and near the AF regime. 
$C_p$ follows the simple $\gamma T + \beta T^3$ law at low temperatures, and electronic 
specific heat coefficient $\gamma$ is found to increase systematically with doping even in 
the AF regime ($0 \leq x < 0.02$). This result suggests that a finite density of states at the Fermi level 
is already created with 0.5\%-hole doping, and that the doped Mott insulator is not a real insulator 
with a clear gap.

\end{document}